\documentclass[conference]{IEEEtran}

\usepackage{url}
\usepackage{cite}
\usepackage{xspace}
\usepackage{amsmath,amssymb,amsthm}
\usepackage[pdftex]{color,graphicx}
\usepackage{adjustbox}

\newcommand{\eg}{\textit{e.g.}\xspace,\xspace}
\newcommand{\etal}{\textit{et~al.}\@\xspace}

\newcommand{\ie}{\textit{i.e.}\xspace,\xspace}

\newcommand{\SIF}{SIF\@\xspace}
\newcommand{\SIFlong}{Serial Interpolation Filter\@\xspace}
\newcommand{\cSIF}{c{\SIF}\@\xspace}

\newcommand{\refsec}[1]{Section~\ref{#1}\@\xspace}
\newcommand{\reffig}[1]{Figure~\ref{#1}\@\xspace}

\newcommand\Mark[1]{\textsuperscript#1}

\IEEEoverridecommandlockouts

\begin{document}



\title{Secure Distributed Membership Tests via Secret Sharing\\
\vspace{0.05in}
\large{\bf{How to Hide Your Hostile Hosts: Harnessing Shamir Secret Sharing}}}

  


\author{
\IEEEauthorblockN{David J. Zage\Mark{1}
\thanks{\Mark{1}Authors listed in reverse alphabetical order to account for a lifetime of being 
alphabetically challenged.}}

\IEEEauthorblockA{Sandia National Laboratories\Mark{2}
\thanks{\Mark{2}
Sandia National Laboratories is a multi-program
laboratory managed and operated by Sandia Corporation, a wholly
owned subsidiary of Lockheed Martin Corporation, for the
U.S. Department of Energy's National Nuclear Security Administration
under contract DE-AC04-94AL85000. Approved for unlimited release: SAND2015-5471 C}\\
Albuquerque NM, 87123\\
djzage@sandia.gov}
\and
\IEEEauthorblockN{Helen Xu, Thomas M. Kroeger,\\Bridger Hahn, Nolan P. Donoghue}
\IEEEauthorblockA{Sandia National Laboratories\\
Livermore, CA, 94550\\
\{npdonog, bhahn, hxu, tmkroeg\}@sandia.gov}
\and
\IEEEauthorblockN{Thomas R. Benson}
\IEEEauthorblockA{Tufts University\\Medford MA, 02155\\
thomas.benson@tufts.edu}
}


\maketitle

\begin{abstract}
Data security and availability for operational use are frequently seen as 
conflicting goals.  Research on searchable encryption and
homomorphic encryption are a start, but they typically build from
encryption methods that, at best, provide protections based on problems
assumed to be computationally hard.  By contrast, data encoding methods such as secret sharing 
provide information-theoretic data protections.  Archives that distribute 
data using secret sharing can provide data protections that are resilient to 
malicious insiders, compromised systems, and untrusted components.

In this paper, we create the \emph{\SIFlong}, a method for storing
and interacting with sets of data that are secured and distributed
using secret sharing. We provide the ability to operate over
set-oriented data distributed across multiple repositories without
exposing the original data. Furthermore, we demonstrate the security
of our method under various attacker models and provide protocol
extensions to handle colluding attackers.  The \SIFlong provides 
information-theoretic protections from a single
attacker and computationally hard protections from colluding
attackers.

\end{abstract}

\section{Introduction}\label{sec:intro}

Computer and data compromises have become so frequent it is almost
clich\'e to cite them as motivation for cyber-security research.
Rather than believing every attack can be prevented, effective organizations
must operate under the expectation of eventual
compromise~\cite{Zaichkowsky2013}. Typical data protection methods
such as symmetric encryption depend
on the enduring privacy of a single key 
and thus can be insufficient for long-term data security.

Distributed repositories using data splitting and encoding techniques have been presented as promising
ways to provide data security even with compromised components
\cite{Ganger2001,Subbiah2005,Storer2009,Kroeger2013}. These systems 
use encodings such as secret
sharing~\cite{Shamir1979,Benaloh1996,Beimel2011} to split the bytes of a file into $N$
shares, any $k$ of which can be used to recreate the original data.  Once the shares have been generated, 
they are then distributed across multiple repositories.  This
approach creates an archive that is resilient to insider threat and is
able to ensure data privacy and integrity with as many as $k-1$
repositories compromised.  Additionally, secret sharing has the benefit 
of being information-theoretically secure, unlike many cryptographic 
methods which are based on problems assumed to be computationally hard.

Most previous work on secret sharing has assumed access to the
stored information will occur through full reassembly of the data.  As a
result, this data would then exist in a single location and be significantly
more vulnerable than when it was split into shares.   For
several use cases, such as set membership, full assembly of the secret may not be necessary.
Thus, performing a full assembly is an unjustifiable security risk. 

To illustrate how such a system would work, we present an example of
five companies who want to share a list of known bad IP addresses.  No
one company trusts any other individual company, but they trust the
group as a whole.  Any company should be able to query and insert IP
addresses, but none should be able to access the entire list.  Using
secret sharing, the companies split, exchange, and store the
addresses.  While the addresses are secure at rest, a mechanism for
query without reconstruction is necessary.

To address this need, we present the \SIFlong (\SIF), which allows
collaborators to store a set of values and support membership queries and inserts
without exposing the original values.  Our contributions include:
\begin{itemize}
\item an information-theoretically secure method for operating on set-oriented data stored across
multiple repositories.
\item a security and performance analysis of our method, demonstrating the ability to 
maintain data confidentiality in noncollusive environments while maintaining performance.
\item an extension to \SIF using the discrete logarithm as a cryptographic trapdoor to mitigate colluding adversaries.
\end{itemize}

The rest of the paper is organized as follows: we review related work
and background information in \refsec{sec:background}, describe the
\SIFlong in \refsec{sec:sif}, analyze our method in \refsec{sec:analysis},
discuss handling collusion in \refsec{sec:byzantine}, and
conclude our work in \refsec{sec:conclusion}.

\section{Background And Related Work}\label{sec:background}

In this section, we present an overview of secret sharing, including recent advances in the area,
and compare our work with searchable encryption.

\subsection{Secret Sharing}\label{ssec:background:christine}

Shamir~\cite{Shamir1979} originally developed secret sharing as an
information-theoretically secure approach to share and store a secret amongst
a group with $N$ members but reconstruct the information with only $k$ of the members.
The algorithm shares a secret amongst multiple participants by selecting a polynomial of at most degree $k-1$, 
setting the $y$-intercept of the polynomial to be the desired secret,
and distributing points on the polynomial with non-zero $x$ values.

To demonstrate how Shamir's algorithm works, suppose Christine has a safe with the combination
$d=854$.  In an emergency, Christine wants to allow any three of the five people
from her office to combine their information to open her safe. She would also
like to prevent any fewer than three people from gaining any information about her combination.
To do so, she encodes the combination, $d=854$, into $N=5$ shares and
requires $k=3$ of these shares to recover the secret.  To do
this, she generates a random polynomial of degree at most $k-1$, since any three
points can uniquely identify this polynomial:
\[
p(x) = d + 276 x + 53 x^2.
\]

\begin{figure}[t]
  \begin{center}

    \begin{adjustbox}{max totalsize={2.0in}{2.0in},center}
    \includegraphics[width=0.9\columnwidth]{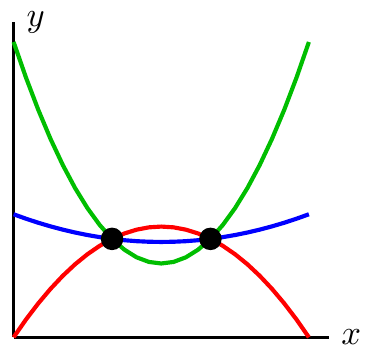}
    \end{adjustbox}

  \end{center}
  \caption{Given two points, all possible values of $p(0)$ are equally likely.}
  \label{infinite_polynomials}
\end{figure}

Christine's safe combination is encoded as the value of a polynomial
curve at $x=0$.  Christine now creates shares by evaluating the polynomial
at $x$-values other than $0$. By combining any three of
these points, Christine's officemates can solve for the original
polynomial and recover the combination. However, with only two
shares, there are as many possible intercepts as there are possible values
for $d$, all equally likely.  Fewer than three shares divulges no
information about the combination. \reffig{infinite_polynomials}
illustrates how, given two specific points, any possible value
of $d=p(0)$ is an equally likely solution.

In practice, we typically perform all secret sharing operations over
a finite field. This allows us to choose polynomial coefficients
randomly over a uniform distribution over the elements of the field
and obviates floating point errors~\cite{Shamir1979}.

This area continues to be an active field of research with many different variations of secret sharing~\cite{Shamir1979,Benaloh1996,Beimel2011,Chor1985,Gennaro1998,Herzberg1995}. 
Notably, Narayanan~\etal discuss information-theoretically secure membership tests for a secret-shared set~\cite{Narayanan2009}. However, the work assumes the set is static and the user issuing queries cannot collude with shareholders.

\subsection{Searchable Encryption}
This work is akin to searchable
encryption~\cite{Song2000} and homomorphic encryption
algorithms~\cite{Gentry2009, Gentry2010}.  Such
algorithms allow a user to perform certain operations such as remote
keyword searches or computing functions on encrypted data on an
untrusted server without decrypting the information or pulling all of
the encrypted data back to the user.  Recent advances in searchable
encryption include multi-keyword ranked searching on cloud data with
low computation~\cite{Cao2014} and work with limited size on a
user's mobile device~\cite{Chang2005}.

However, the fundamental security protections provided by encrypted
data have many limitations. An attacker that compromises the single
host gains access to all of the data.  Solutions to searching
encrypted data are computationally secure rather than
information-theoretically secure and depend on the assumed hardness of
certain problems.  Homomorphic encryption is typically very expensive,
and while specialized hardware exists, it is limited in utility.  Even
encryption with multiple hosts does not capture the benefits of secret
sharing because the data is replicated instead of split into shares.
To the best of our knowledge there has been little work on operating
on data in a secret-shared archive.

\section{Serial Interpolation Filter}\label{sec:sif}

In this section, we provide a detailed description of the \SIFlong
method.  We show this method can enable operational use of the data
while maintaining information-theoretic data security, a protection
well above commonly used symmetric encryption techniques.

\subsection{Using Secret Sharing to Store a Set}
\label{ssec:background:set}

We draw from previous work~\cite{Ganger2001,Subbiah2005,Storer2009,Kroeger2013}
that proposed using secret sharing to create a secure, distributed
archive.  While the archive is usable for any type of data representable as a set, 
we use IP addresses as our exemplar.

\begin{figure}[t]
\begin{center}
\begin{adjustbox}{max totalsize={0.95\columnwidth}{0.95\columnwidth},center}
\includegraphics[width=0.95\columnwidth]{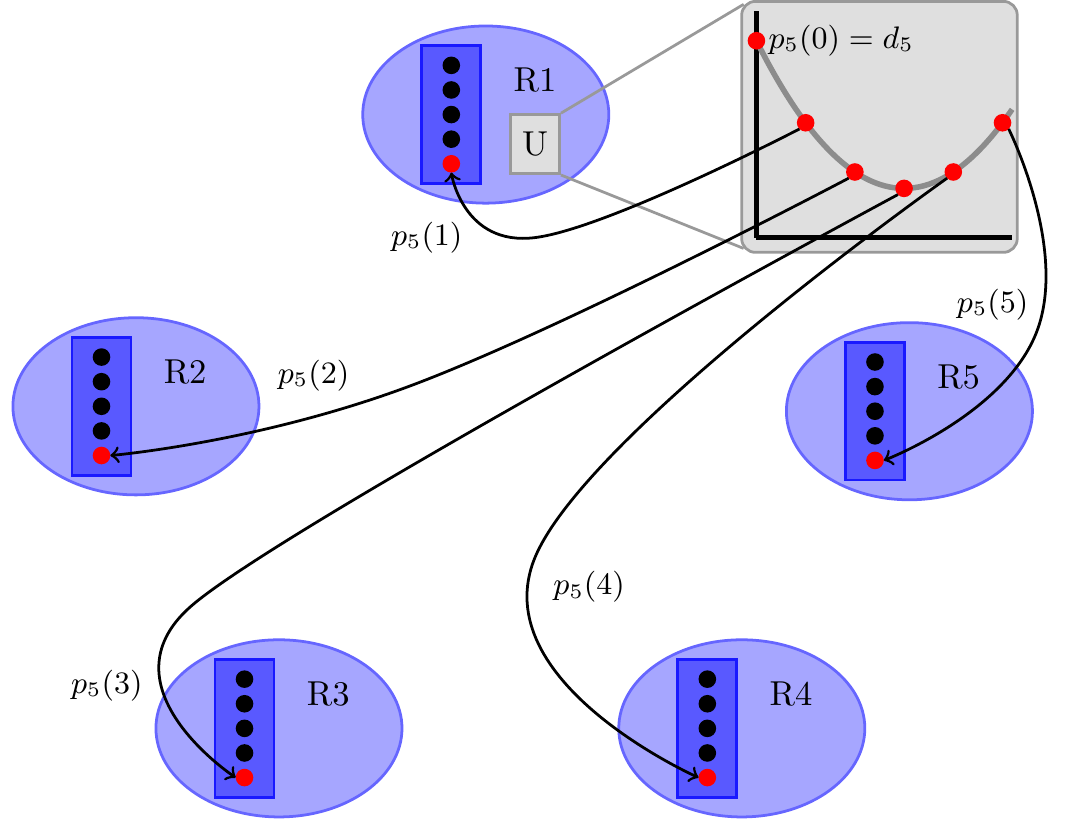}
\end{adjustbox}
 \end{center}
\caption{An example of share distribution to securely store an entry in a distributed archive.}
\label{fig:createList}
\end{figure}

A trusted user creates a new entry after externally determining a specific
IP is untrusted.  To securely store this element, it is split into $N$
shares using a $(k,N)$ linear secret sharing scheme~\cite{Shamir1979}.
As shown in \reffig{fig:createList}, once the shares are created, they
are then distributed across $N$ separate repositories to create the
archive holding the list of known bad IP addresses.  For example,
if there are 20 bad addresses, each of the $N$ repositories would hold
20 shares unique to that repository.  The aggregate of these repositories
is an archive securely
storing 20 addresses. Once the shares have been distributed, the
user deletes the original data.  After this, the list exists solely as an abstract
concept and no single repository holds data recognizable as an
IP address (or component thereof) in isolation.  It is this unique
property that separates such solutions from systems using encryption
for data security.

When one of the repositories is inevitably compromised~\cite{Zaichkowsky2013}, 
the attacker learns nothing beyond the size of the set of elements.  Even additional 
compromises, as long as they number less than $k$, do not let the attacker 
gain any additional information about the list contents. Ideally, the 
consortium is able to refresh the shares with new polynomials using 
techniques such as those by Herzberg~\etal~\cite{Herzberg1995} before the 
intruders have enough shares to reconstruct the data.

\subsection{Terminology}\label{ssec:sif:terminology}
We define the following terms:

\begin{itemize}
\item {\bf Element}---One entry in the set of data. We wish to simultaneously
protect and use these elements.
\item {\bf Share}---A resulting datum from using secret sharing to encode an
  element into $N$ pieces.
\item {\bf Threshold}--- The number of shares, denoted as $k$, needed to recover the original
  element.
\item {\bf Repository}---A remote server storing a unique set of shares.
\item {\bf Archive}---The collection of all $N$ repositories. Together they
  create a system for securely storing and operating on data elements.
\end{itemize}

We perform all secret sharing operations over a finite field $F$.
The data is a set $D = \{d_\ell | d_\ell\in F, 1 \leq \ell\}$ and is of size $|D|$.
We split each data element $d_\ell \in D$ using a polynomial,
$p_\ell(x)$, of order $k-1$.  The polynomial takes the form
\[
p_{\ell}(x) = d_{\ell} + a_{\ell, 1} x + \dots + a_{\ell, k -1} x^{k-1}
\]
where $a_{\ell, i} \in F$ is a coefficient chosen uniformly at random for
$\ell\in\{1,\ldots,|D|\}, i \in \{1,\ldots,k-1\}$. 

We denote the list of shares given to Repository $r$ as
$\vec{p}(x_r)$, which is the vector of all of the polynomials
evaluated at $x_r$.  We use the vector notation to represent
interpolation performed on all shares concurrently.

\subsection{Serial Lagrangian Interpolation}\label{ssec:sif:sli}

In traditional secret sharing,
given a set of $k$ points, $S= \{x_1,\dots,x_k\}$ and their
corresponding shares ($p(x_r)$ for $x_r \in S$), we can use
Lagrangian interpolation to reconstruct the generating polynomial,
$p(x)$ as

\[
p(x) = \sum_{i=1}^{k}L_{i,S}(x)p(x_i),
\]
where
\[
L_{i,S}(x) = \!\prod\limits_{\substack{x_j \in S \\ x_j \neq x_i}}\! \frac{x-x_j}{x_i-x_j}.
\]

This allows for recovery of the original element by evaluating
$p(0)$.

The archive stores the elements as the $y$-intercepts of polynomials,
$\vec{d}=\vec{p}(0)$.  Here, we use the vector notation to denote operation 
over all elements in the set. In our example we
select a finite field able to represent $2^{32}$.

To avoid reconstructing the polynomials in a single location, we
perform Lagrangian interpolation serially across a subset of $k$ out
of $N$ repositories.  The list of $k$ repositories to be used is
provided as a part of the \SIF query and known to all of the
repositories involved.  Furthermore, the data of each repository,
$\vec{p}(x_r)$, is private and known only to the repository owning
that data. Therefore, while all of the repositories can compute any of
the $L_{i, S}(x)$, only Repository $r$ has access to $\vec{p
}(x_r)$. Thus it is possible to calculate $\vec{p}(0)$ across three
repositories ($k=3$) as follows:

\[
\vec{p}(0) = \underbrace{L_{1, S}(0) \vec{p}(x_1)}_{\text{Repository 1}} +
 \underbrace{L_{2,S}(0) \vec{p}(x_2)}_{\text{Repository 2}} +
 \underbrace{L_{3,S}(0) \vec{p}(x_3)}_{\text{Repository 3}}.
\]

\begin{multline*}
\vec{p}(0) = \underbrace{\frac{x_2 x_3}{(x_1-x_2)(x_1-x_3)} \vec{p}(x_1)}_{\text{Repository 1}}\\
+ \underbrace{\frac{x_1 x_3}{(x_2-x_1)(x_2-x_3)} \vec{p}(x_2)}_{\text{Repository 2}}\\
+ \underbrace{\frac{x_1 x_2}{(x_3-x_1)(x_3-x_2)} \vec{p}(x_3)}_{\text{Repository 3}}.
\end{multline*}

Nevertheless this calculation would reassemble the original secret in
a single location. To perform a set membership test while not exposing the
polynomial or query term, we must perturb both the query and the
polynomial reconstruction with a vector of nonces, or randomly generated
constants. This process is described in 
detail in the next section.

\subsection{Generalized Algorithm}\label{ssec:sif:general}

\begin{figure}[t]
\begin{center}
\begin{adjustbox}{max totalsize={0.95\columnwidth}{0.95\columnwidth},center}
\includegraphics[width=0.95\columnwidth]{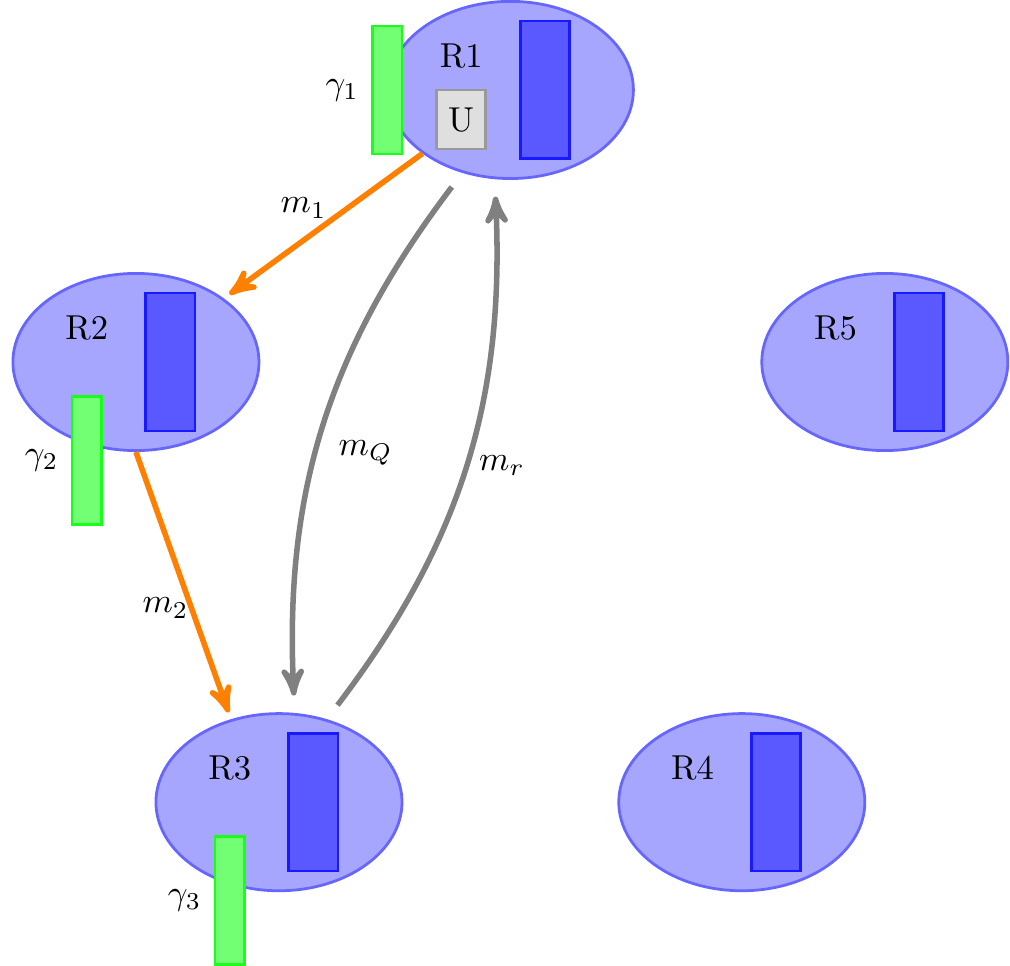}
\end{adjustbox}
\end{center}
\caption{An example of query algorithm for testing set membership.}
\label{fig:query}
\end{figure}

Once the shares have been distributed, we can now use the \SIF to
test for set membership. We present
the query algorithm a user at a repository would perform to determine if
the value $Z$ is present in the archive.  An example query round and 
user response for a $(k=3,N=5)$ secret sharing can be seen in 
\reffig{fig:query}.

\begin{enumerate}
\item\label{enum:assemb:one} A user at repository $R_1$ initiates the
  query with a unique transaction identifier $q$.   $R_1$ then
  selects $k$ repositories including itself and creates an ordered list of these
  repositories $S = [x_1, \dots, x_k]$.  It generates a nonce
  vector ($\vec{\nu}$) containing a different random pad for each element
	and calculates $\vec{\gamma}_1$ as
  follows: \[ \vec{\gamma}_1 = L_{1, S}(0) \vec{p}(x_1) + \vec{\nu} \]

\item\label{enum:assemb:two}  $R_1$ sends a membership test
  message ($m_1 = [q, \vec{\gamma}_1, S]$) to the next repository listed in $S$.
	In \reffig{fig:query}, $R_1$ sends the message to $R_2$.

\item\label{enum:assemb:three}  $R_1$ calculates a nonced query
  term $\vec{Q}=Z\vec{1}+\vec{\nu}$ and sends a separate message, $m_Q = [q,\vec{Q}]$, to the $k^{\text{th}}$ repository.
	As $k=3$ in \reffig{fig:query}, this message is sent from $R_1$ to $R_3$.



\item\label{enum:assemb:five}  $R_i$ for $i\in\{2, \ldots, k-1\}$ receives the message from $R_{i-1}$ and
   serially calculates its portion of  the interpolation
   polynomials
   \begin{equation*}
        \label{eqn:gammai}
                 \vec{\gamma}_i =  L_{i, S}(0) \vec{p}(x_i) + \vec{\gamma}_{i-1}
   \end{equation*}
   and sends $m_i=[q, \vec{\gamma}_i, S]$ to repository $R_{i+1}$.

\item\label{enum:assemb:six}  $R_k$: Calculates the final
  contribution to the interpolations
  \begin{equation*}
    \label{eqn:gammak}
    \vec{\gamma}_k = L_{k,S}(0)   \vec{p}(x_k) + \vec{\gamma}_{k-1}
  \end{equation*}
  and compares each component in $\vec{\gamma}_k$ to each corresponding
  component of the query terms ($\vec{Q}$) received in message $m_Q$ in step~\ref{enum:assemb:three}. 
	If a match is found, the query response is set to true and false otherwise.

\item\label{enum:assemb:seven}  $R_k$ sends the results of the query to the 
	request originator $R_1$ ($m_r=[q,True|False]$). This final response is 
	seen as the gray arrow returning from $R_3$ to the user at $R_1$ in \reffig{fig:query}.
\end{enumerate}

The user $U$ can determine if the value $Z$
is in the set while no other user learns the value of $Z$. Additionally,
none of the secrets ($\vec{d}=\vec{p}(0)$) are ever reconstructed in a
single location. 
In short, the \SIF protocol has
enabled $U$ to actively query the data while maintaining
information-theoretic levels of data
protection. In \refsec{sec:analysis}, we will discuss these
protections in detail and explain some limitations around adversarial
collusion.

\section{Analysis}\label{sec:analysis}
In this section, we present the threat models we will use for subsequent security and performance analyses.  

\subsection{Threat Model}\label{ssec:analysis:model}

In analyzing the \SIFlong, we focus on the confidentiality of the
stored data and minimizing release of information due to legitimate
user queries. We assume users are authenticated to the service and all
legitimate users have the right to place queries, which results in the
user learning if the item was contained (or not) in the set, but
learning nothing else about the set.  We assume the 
presence of secure communications channels.

With this in mind, we consider two adversary models:
\begin{itemize}
	\item Honest-but-curious participants
	\item Byzantine participants
\end{itemize}
We start with the most restrictive model from the attacker's standpoint and gradually relax these assumptions while strengthening our protocol.  
In the honest-but-curious model, the attacker (\ie any participating party including the end-user) must correctly
follow all parts of the protocol, which includes sending correct protocol responses.  The attacker is allowed to perform extra calculations
and store previous protocol values, but is not allowed to actively aggregate information it would not normally have received during protocol
participation (\ie no collusion).  Relaxing our assumptions under a Byzantine model, we consider a system containing $N$ participants and a 
bounded number of malicious nodes $0 \leq f < k$, where $k$  is the threshold value for the scheme. 
The malicious nodes behave 
arbitrarily and are only limited by the constraints of any cryptographic methods deployed \cite{Dolev1983}, which are assumed resistant to 
tampering.  The set of malicious nodes may collude.  

As we focus on the confidentiality of the data and the privacy of the queries, attacks targeting the integrity (\eg reporting incorrect shares)
or availability (\eg withholding shares) of the secret sharing algorithm are outside the scope of this paper.  Techniques such as proactive,
public, or verifiable secret sharing \cite{Chor1985,Gennaro1998,Herzberg1995,Beimel2011} can be used to augment our solution to alleviate many of 
these issues, but are left for future work.

\subsection{Security}\label{ssec:analysis:sec}
Fundamentally, the security of the data rests on the privacy of
each repository's shares, $\vec{p}(x_i)$.  It is the ability to aggregate or otherwise
calculate these values that constitutes a loss of data
protections.  We also note that given fewer than $k$ shares, as proved
by Shamir and others, an attacker gains nothing.  Intuitively, this is the same
as defining a specific parabola given only two points as shown
in~\reffig{infinite_polynomials}(\ie there are as many possible coefficients as
there are distinct elements in the field).

For the honest-but-curious model, the repositories are
unable to calculate any specific shares held by other
repositories.  For $R_2$ through $R_k$, $\vec{\gamma}_i$ is additively
perturbed by $\vec{\nu}$.  Assuming strong random number generation,
this provides the same protections as a one-time pad for all
repositories except $R_1$. Since $R_1$ only receives a true/false
result it has no further information about the shares at any of the other
repositories.  The result is our system maintains information-theoretic 
protections on data confidentiality.

Under our Byzantine model, $\vec{\gamma}_k$ is protected only by
$\vec{\nu}$ and malicious repositories may share partial results. Thus, 
collusion between $R_1$ who holds $\vec{\nu}$ and
$R_k$ would provide access to the complete set of secrets.  While this
vulnerability to collusion provides data exposure, we note this is not the
case for the data at rest and we still provide data protections 
better than those of typical encryption-based methods.
Nevertheless, in~\refsec{sec:byzantine} we present a method using
discrete logarithms to provide computationally hard protections against
collusion.

\subsection{Performance}\label{ssec:analysis:perform}

Recall from the generalized SIF algorithm that $k$ denotes the number of 
repositories involved in one query round. Furthermore, we use $|D|$ to denote the 
number of data elements in the list. We assume it takes $O(1)$ to compute a nonce and to send a single data element.
It takes $O(k)$ work to evaluate the Lagrange polynomial at each of the $k$ repositories
in the round, resulting in $O(k^2 + |D|)$ work to carry out the entire interpolation.

Additionally, each message passed contains $O(|D|)$ data elements and $O(k)$ repository labels and
therefore takes $O(|D| + k)$ work per message. Therefore, the work for message transfers in a given
round of the protocol is $O(|D|k +k^2)$.

The total work includes both the transfers and interpolation giving a final result
of $O(|D|k + k^2)$. We generally expect $|D|$\textgreater\textgreater$k$, that is,
the number of data elements vastly exceeds the number of repositories
necessary to carry out a query. The computation is dominated by the messages sent
rather than the calculations done at each repository.

Similar arguments can be made for insertion using $O(N)$ messages and $O(N)$ computation.

\section{Mitigating Byzantine Adversaries}\label{sec:byzantine}

As was demonstrated in \refsec{ssec:analysis:sec}, the SIF
protocol is resilient to honest-but-curious adversaries but does not
hold these guarantees in the face of adversarial collusion.  In this
section, we present a method for maintaining security in Byzantine
environments based on computational guarantees.  By utilizing a
cryptographic trapdoor function based on the discrete logarithm
problem\cite{Stinson2005}, we are able to create computational \SIF (\cSIF)
that is secure given at most $k-1$ adversaries. Consider a (multiplicative) 
cyclic group $C_q$ of order $q$ with
generator $g$ where the discrete logarithm problem is assumed hard \cite{Odlyzko1985}.
Under this scenario, we now present the
\cSIF query algorithm a user at a repository would perform to determine if
the value $Z$ is present in the archive.  Original share distribution and storage remains unchanged.

We present some notation regarding vectors. $g^{\vec{v}}$ represents a vector
where the $\ell^{\text{th}}$ component is $g^{v_\ell}$, the generator 
raised to the power of the $\ell^{\text{th}}$ component of $\vec{v}$.
The operator $\odot$ denotes component-wise multiplication. For example, each component $c_i$ 
of $\vec{c}=\vec{a} \odot \vec{b}$ is defined as $c_i = a_ib_i$.
\begin{enumerate}
\item\label{enum:byz:one} A user at repository $R_1$ initiates the
  query with a unique transaction identifier $q$.   $R_1$ then
  selects $k$ repositories including itself and creates an ordered list of these
  repositories $S = [x_1, \dots, x_k]$.  It then generates a nonce
  vector ($\vec{\nu}$) and calculates $\vec{\gamma}_1$ as
  follows: \[ \vec{\gamma}_1 = g^{L_{1, S}(0) \vec{p}(x_1) + \vec{\nu}} \]

\item\label{enum:byz:two}  $R_1$ sends a membership test
  message ($m_1 = [q, \vec{\gamma}_1, S]$) to the next repository listed in $S$ ($R_2$).

\item\label{enum:byz:three}  $R_1$ calculates a nonced query
  term $\vec{Q}=g^{Z\vec{1}+\vec{\nu}}$ and sends a separate message to the $k^{th}$ repository 
  ($m_Q = [q,\vec{Q}]$).

\item\label{enum:byz:five}  $R_i$ for $i\in\{2, \ldots, k-1\}$ receives the message from $R_{i-1}$ and
   serially calculates its portion of  the interpolation
   polynomial
   \begin{equation*}
        \label{eqn:gammaib}
                 \vec{\gamma}_i = \vec{\gamma}_{i-1} \odot g^{ L_{i, S}(0) \vec{p}(x_i)} 
   \end{equation*}
   and sends $m_i=[q, \vec{\gamma}_i, S]$ to repository $R_{i+1}$.

\item\label{enum:byz:six}  $R_k$: Calculates the final
  contribution to the interpolation
  \begin{equation*}
    \label{eqn:gammakb}
    \vec{\gamma}_k = \vec{\gamma}_{k-1} \odot g^{L_{k,S}(0)  \vec{p}(x_k)}
  \end{equation*}
  and compares each component in $\vec{\gamma}_k$ to each corresponding
  component of the query terms ($\vec{Q}$) received in step~\ref{enum:byz:three}. 
	If a match is found, the query response is set to true and false otherwise.

\item\label{enum:byz:seven}  $R_k$ sends the results of the query to the 
request originator $R_1$ ($m_r=[q,True|False]$).
\end{enumerate}

\subsection{Security Sketch}\label{ssec:byz:sec}
The constituents of $\vec{\gamma}_i$ for $i \in \{1,\cdots,k\}$ are now protected as
exponents of the generator function (\eg $g^{L_{i, S}(0) \vec{p}(x_i)}$).

Assuming the discrete logarithm is hard, the colluding repositories cannot compute the logarithm of the 
 messages and find the original share values.  This would be equivalent to solving the Computational Diffie-Hellman problem \cite{Boneh1998}.  
Thus, even under the Byzantine model which allows for collusion between $R_1$ who holds $\vec{\nu}$ and
$R_k$ who holds $\vec{\gamma}_k$, the two repositories cannot gain any additional stored secrets.  This holds true for 
any subset of $k-1$ colluding repositories which do not have enough shares to recreate the original data. 
The primary use of the nonce vector in this case is to blind query messages and responses.

\section{Conclusions and Future Work}\label{sec:conclusion}

Modern systems desperately need a way of securing data in spite of 
compromise.  While techniques like Shamir's Secret Sharing have been around 
for decades, the ability to operationally use data without exposing the 
original information provides an important capability in efforts to ensure 
secure and resilient computer systems.  Our \emph{\SIFlong} uses the strong 
data protections from secret sharing to secure data at rest.  In addition, we 
enable active use of data while maintaining information-theoretic levels of 
protection during a query. Although collusion between two members can expose 
the original data set, we show how a cryptographic trapdoor based on discrete 
logarithms can be used to enable computationally hard resilience to up to $k-1$ 
colluding adversaries.

Future work includes minimizing the amount of data the SIF protocol must 
store and transfer using advanced data structures such as a quotient filters 
\cite{Bender2012} or cuckoo filters\cite{Fan2014}. Additionally, while the 
\cSIF algorithm was able to tolerate collusion, we hope to extend \SIF 
in a manner tolerating collusion among attackers while maintaining 
information-theoretic security guarantees.

\section*{Acknowledgments}
\addcontentsline{toc}{section}{Acknowledgments}\label{sec:acks}
We would like to thank Cindy Phillips, Jonathan Berry, Michael Bender, Rob Johnson, and Jared Saia for their insights and thoughtful discussions. This work was supported by the Laboratory Directed Research and Development Program at Sandia National Laboratories.

\bibliographystyle{IEEEtran}
\bibliography{icnc-2016}

\end{document}